\documentclass[10pt,journal,compsoc]{IEEEtran}

\usepackage[switch]{lineno}
\ifCLASSOPTIONcompsoc
  \usepackage[nocompress]{cite}
\else
  \usepackage{cite}
\fi

\ifCLASSINFOpdf
\else
\fi

\AtBeginDocument{%
  \providecommand\BibTeX{{%
    \normalfont B\kern-0.5em{\scshape i\kern-0.25em b}\kern-0.8em\TeX}}}
    
\usepackage{url}
\usepackage{multirow}
\usepackage{array}
\usepackage{tikz}
\usepackage{sparklines}
\usepackage{subfig}
\usepackage{graphicx}
\usepackage{makecell}
\usepackage[utf8]{inputenc}
\usepackage{tcolorbox}

\usepackage{amsmath}
\usepackage{stfloats}
\newtcbox{\inlinecode}{on line, boxrule=0pt, boxsep=0pt, top=2pt, left=2pt, bottom=2pt, right=2pt, colback=gray!15, colframe=white, fontupper={\ttfamily \footnotesize}}
\usepackage{xcolor}
\usepackage{pgf-pie}
 
\usepackage[inline]{enumitem}
 \usepackage{algorithm}
 \usepackage{algorithmic}
\usetikzlibrary{arrows,positioning,shapes.geometric}
\usepackage{tcolorbox}
\begin{document}
\title{\Large \bf Are your dependencies code reviewed?: \\Measuring code review coverage in dependency updates}

\author{Nasif~Imtiaz
        and~Laurie~Williams% <-this % stops a space
\IEEEcompsocitemizethanks{\IEEEcompsocthanksitem Nasif Imtiaz and Laurie Williams are with the Department of Computer Science, North Carolina State University, Raleigh, NC, 27606.\protect\\
E-mail: {simtiaz, lawilli3}@ncsu.edu}% <-this % stops an unwanted space
% \thanks{Manuscript received April 19, 2005; revised August 26, 2015.}
}
\IEEEtitleabstractindextext{%
\begin{abstract}
As modern software extensively uses free open source packages as dependencies, developers have to regularly pull in new third-party code through frequent updates. However, without a proper review of every incoming change, vulnerable and malicious code can sneak into the codebase through these dependencies.
The goal of this study is to aid developers in securely accepting dependency updates by measuring if the code changes in an update have passed through a code review process.
We implement Depdive, an update audit tool for packages in Crates.io, npm, PyPI, and RubyGems registry. Depdive first (i) identifies the files and the code changes in an update that cannot be traced back to the package's source repository, i.e., \textit{phantom artifacts}; and then (ii) measures what portion of changes in the update, excluding the phantom artifacts, has passed through a code review process, i.e., \textit{code review coverage}.

Using Depdive, we present an empirical study across the latest ten updates of the most downloaded 1000 packages in each of the four registries. We further evaluated our results through a maintainer agreement survey. We find that phantom artifacts are not uncommon in the updates (20.1\% of the analyzed updates had at least one phantom file). The phantoms can appear either due to legitimate reasons, such as in the case of programmatically generated files, or from accidental inclusion, such as in the case of files that are ignored in the repository. 
% However, without provenance tracking, we cannot audit if the changes in these phantom artifacts were code-reviewed or not.
Regarding code review coverage (\textit{CRC)}, we find 
the updates are typically only partially code-reviewed (52.5\% of the time). Further, only 9.0\% of the packages had all their updates in our data set fully code-reviewed, indicating that even the most used packages can introduce non-reviewed code in the software supply chain. We also observe that updates either tend to have high \textit{CRC} or low \textit{CRC}, suggesting that packages at the opposite end of the spectrum may require a separate set of treatments.
\end{abstract}
}
\maketitle

\section{Introduction}
Modern software extensively uses free open source packages as \textit{dependencies}~\cite{blackduck2021}. However, using open source has opened up new attack vectors, as vulnerable and even malicious code can sneak into software through these third-party dependencies~\cite{ohm2020backstabber}. Further, practitioners are recommended to keep dependencies up to date with the latest version~\cite{whykeepupdated}, resulting in developers pulling in new code through frequent  updates~\cite{monthlyupdate}, often automatically and without a security review~\cite{mirhosseini2017can}.

Recent times have seen popular open source packages be compromised and push malicious updates, such as the attack through \textit{ua-parser-js}, an npm package used by major software corporations~\cite{uaparserjs}. Attacks through dependencies, such as the preceding example, are categorized as ``supply chain attacks'' by security practitioners~\cite{deming2020good}. While the existence of malicious packages in different ecosystems is well-known~\cite{ohm2020backstabber}, popular packages from reliable sources can also be compromised, such as through:
\begin{enumerate*}
    \item hijacking of a maintainer's account~\cite{usparserjshicack};
    
    % ~\footnote{Attacker(s) hijacked the npm account of \textit{ua-parser-js} maintainer~\cite{usparserjshicack}.}
    \item a maintainer going rogue~\cite{roguemaintainer};
    % ~\footnote{The maintainer of npm packages \textit{colors} and \textit{faker} intentionally added infinite loops in new updates~\cite{roguemaintainer}.}
    \item account handover through social engineering~\cite{socialeng}; and
    % ~\footnote{The ownership of npm package \textit{event-stream} was transferred to a malicious user who gained the trust of original owner by making a series of accepted contributions~\cite{socialeng}.},
    \item build system compromise~\cite{travisci}.
    % ~\footnote{Travis CI vulnerability exposed secrets of many open source projects~\cite{travisci}.}, 
\end{enumerate*}

Therefore, developers are now recommended to review dependency updates before merging them into the codebase~\cite{yang2021solarwinds, githubdepreview}, as the responsibility of security lies on the consumer when using free open-source code~\cite{freeliability}. However, manually reviewing each update's code changes may not be a practical solution, as projects may have hundreds of direct and transitive dependencies ~\cite{imtiaz2021comparative, blackduck2021}. Further, actively maintained packages get frequent updates, over-burdening any project that would employ such strict measures. Therefore, we propose that dependency updates go through automated security and quality checks before being merged, which can act as the first line of defense and aid developers in securely accepting dependency updates.

Recently, the software industry has proposed frameworks to define the compliance standards for using open source packages, such as the ``Supply chain Levels for Software Artifacts (SLSA)''~\cite{slsa} framework. Specifically, SLSA provides a checklist of standards and controls to prevent tampering, improve integrity, and secure packages and infrastructure. Among others, SLSA requires a dependency package to employ a two-person code review~\cite{slsareq}. However, while projects like the ``Security Scorecards''~\cite{ossfscorecard} exist to aid in dependency selection, little research has been done yet on employing automated checks during each subsequent update~\cite{vu2021lastpymile}. The goal of this study is to aid developers in securely accepting dependency updates by measuring if the code changes in an update have passed through a code review process.

The check for code review ensures that each line of code change in the update can be traced back to at least two owners. This check can guard against the threat scenarios where \begin{enumerate*}
    \item only one maintainer's account has been hijacked; 
    \item a single maintainer has gone rogue; or
    \item attackers have compromised the publishing infrastructure to inject unowned malicious code.
\end{enumerate*} 
Further, research has shown that code review helps improve code quality~\cite{mcintosh2016empirical} and can prevent the introduction of new security vulnerabilities~\cite{bosu2014identifying}.
Ladisa et al.~\cite{ladisa2022taxonomy} have developed an attack taxonomy for open source supply chain attacks, where they have mentioned \textit{code review} as a safeguard against the attack vector \textit{inject into sources of legitimate package}.
% The risk of vulnerabilities in the dependencies has been extensively studied in the literature~\cite{zimmermann2019small, imtiaz2021comparative}, and even the legitimate updates can bring in new vulnerabilities. 

% and has recently been exemplified by the critical vulnerability discovered in \textit{log4j}, a widely used Java package~\cite{log4shell}. 

However, the feasibility of employing an automated check for the code review requirement during dependency updates has not been studied yet. Packages get bundled and distributed in various ways and may use different code review tooling. Such differences in the maintenance and distribution of packages create challenges in reliably auditing the updates. Further, no empirical study exists on the code review practices among top packages to understand the practicality of this SLSA requirement. This paper aims to address these gaps in the research of securely using open source dependencies.

We implement Depdive which measures the \textit{\textbf{code review coverage (CRC)}}~\cite{mcintosh2014impact} in a dependency update. We define \textit{CRC} as the proportion of the code changes in an update that has gone through a code review process. Depdive works for four package registries, namely Crates.io for Rust, npm for JavaScript, PyPI for Python, and RubyGems for Ruby. We choose these registries because they follow a similar package distribution model, where the maintainers upload the package code to the registry.
We further scope our implementation to GitHub repositories as we leverage the platform to determine if certain code changes have been reviewed or not.
% ~\footnote{Among other package registries, Maven for Java and NuGet for DotNet contain the compiled binary of a package, rather than the source code itself. On the other hand, Go package manager and Composer for PHP do not host their own package repository but act as an intermediary tool to retrieve source code directly from a package's own source code repository.}.

%  We ask the below research questions:
%     \begin{quote}
%         \textbf{RQ1: Can we effectively measure the code review coverage in a dependency update?}
%     \end{quote}

    % \begin{quote}
    %     \textbf{RQ2: What is the code review coverage in the version updates of the most downloaded open source packages?}
    % \end{quote}

In summary, Depdive maps code changes between two versions of a package uploaded in the registry to the corresponding commits in the package's source repository. It then identifies if there was a reviewer for the mapped commits through four GitHub-based checks. Along the process, Depdive also identifies the files and lines of code that cannot be mapped from the registry to the repository, which we refer to as \textit{\textbf{phantom artifacts}}, following the definition in prior work~\cite{vu2021lastpymile}. While one approach can be to consider the phantom artifacts as non-reviewed in the denominator of the \textit{CRC} measurement, these artifacts may exist in the form of binaries or programmatically-generated files, as will be shown in this paper. Therefore, they will require a provenance tracking mechanism to audit if changes in them were reviewed or not. With Depdive, we first filter out the phantom artifacts and output them separately from the \textit{CRC} measurement.

For an empirical evaluation, we run Depdive over the latest ten releases of the most downloaded 1000 packages in Crates.io, npm, PyPI, and RubyGems. Based on Depdive's output, we answer the following two research questions:
\begin{itemize}
    \item \textit{\textbf{RQ1:}} To what extent do phantom artifacts exist in the updates of the most downloaded packages?
    \item \textit{\textbf{RQ2:}} Excluding phantom artifacts, what is the code review coverage (\textit{CRC}) in the updates of the most downloaded packages?
\end{itemize}

Besides the answer to the above two research questions, the contributions of our work include a working tool, Depdive, that outputs details on the phantom artifacts and the code review data for a dependency update. We also present a survey of package maintainers' agreement with Depdive's outputs. The code, survey questionnaire, and data for this paper are anonymously available at \url{https://tinyurl.com/depdive}.

The rest of the paper is structured as follows:
Section \ref{sec:terminology} explains key concepts to this study, Section \ref{depdive} explains Depdive workflow. Section \ref{sec:dataset} explains our dataset, while Section \ref{sec:empirical} presents our findings. Section \ref{survey} presents the maintainer agreement survey to our analysis. Section \ref{limitation} lists the limitation of our study, and Section \ref{sec:discussion} discusses the implication of our findings. Section \ref{relwork} discusses related work before Section \ref{conclusion} concluding this paper.

\section{Background}
\label{sec:terminology}
% Git, Mercurial, and Subversion are the popular tools for version controlling in a repository, while GitHub~\cite{GitHub} and GitLab~\cite{gitlab} are some of the most popular code hosting platforms.
%  in a registry-specific distribution bundle (e.g., \textit{wheel} for PyPI).

% ~\todo{Difference between package registry and repositroy}
% \textit{\textbf{Phantom Artifacts:}}
% Vu et al.~\cite{vu2021lastpymile} have defined a phantom artifact as a software entity, such as files and lines of code, that exist in a package in the registry, but does not exist in the source repository. By definition, the development history of the phantom artifacts is not available in the repository, without a mechanism to trace the origin of these artifacts. 

% A project may be configured to have required policies on which code changes can be merged into the main codebase, e.g., everybody's code changes should go through a code review process or only contributors' code changes should be reviewed by a maintainer of the project. 

% \textbf{\textit{Code Review Coverage (CRC):}} We define code review coverage~\cite{mcintosh2014impact} as the proportion of the code changes in a dependency update that has gone through a code-review process.

In this section, we explain the concepts key to our study:

\textit{\textbf{Package:}} A package, also referred to as a library or module, is a reusable software application unit that can be used by other software.

\textit{\textbf{Dependency:}} When a software uses a package, the package is referred to as a dependency. Moreover, the dependencies of the dependency package itself also become transitive dependencies of the client software. The client software depends on a specific dependency version and can update to a newer version when available.

\textit{\textbf{Package Source Code Repository:}} Repository is a cloud file hosting service with a versioning system to store the source code of a software project. In a git repository, the most granular unit to track a revision of the source code is called a \textit{commit} that is identified through a unique \textit{commit hash}.

\textit{\textbf{Package Registry:}} Package registries are centralized package hosting services to store and distribute packages. In this paper, we work with Crates.io, npm, PyPI, and RubyGems registry. These registries follow a similar package distribution model, where developers can upload their package source code to the registry alongside required artifacts, such as data files and pre-compiled binaries. The client software can then download and install packages from these registries. While it is expected that the same source code in the repository is distributed via the registry, the registry contains its own copy of the package code, which may not be identical to the one in the repository.

\textit{\textbf{Code Review:}} Code review is a manual review process of code changes by any developer(s) other than the author. While the history of code review is not tracked by git, reviews are generally performed using a tool, e.g., Gerrit~\cite{gerrit}. GitHub offers a pull-based development model that integrates native code review tooling. A developer can open a \textit{pull request (PR)} on GitHub to submit code changes and ask for reviews from other developers.

\textbf{\textit{Emerging Industry Standards:}} Recently, multiple industry standards have emerged that provide a check-list of standards and controls for safe use of open source packages. Notable examples are Supply chain Levels for Software Artifacts (SLSA)~\footnote{\url{https://slsa.dev/}} and 
Supply Chain Integrity, Transparency, and Trust~\footnote{\url{https://github.com/ietf-scitt}}.

\section{Depdive}
\label{depdive}
\begin{figure*}
    \centering
    \includegraphics[scale=0.5]{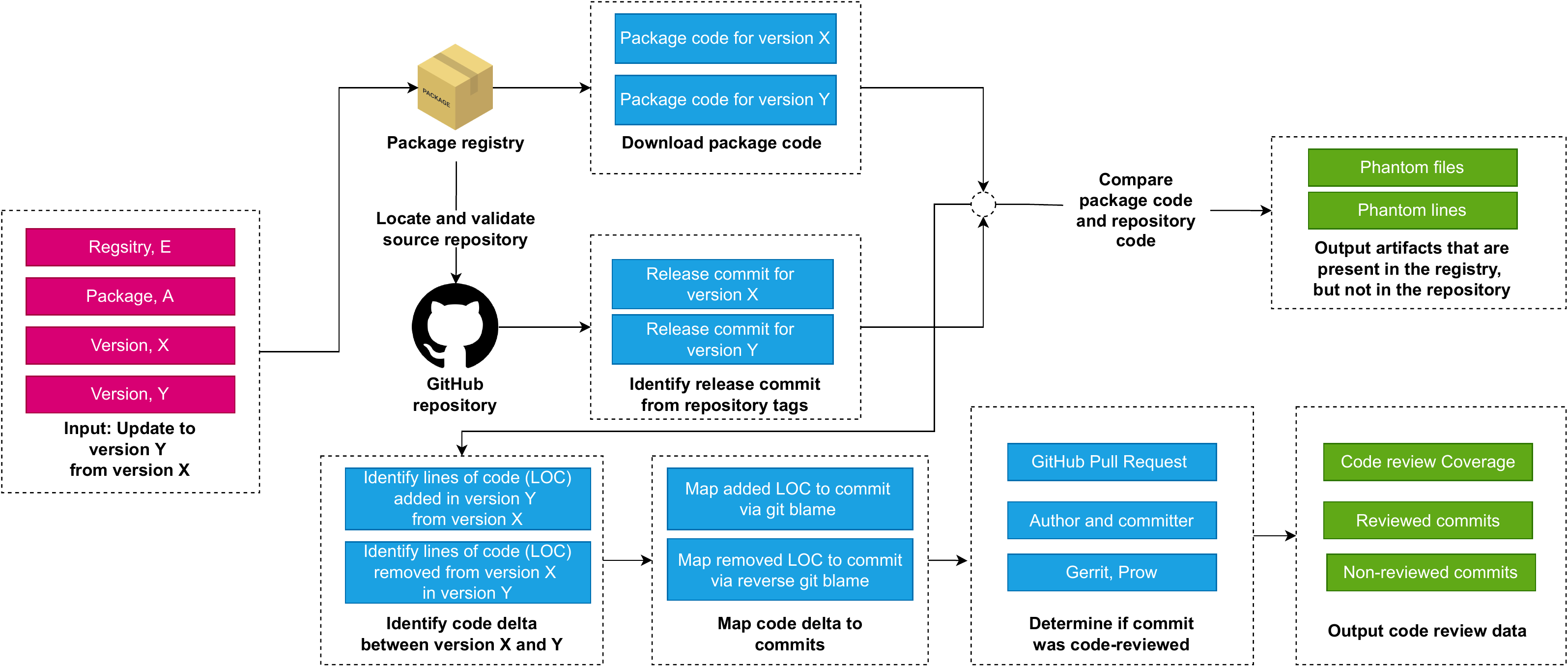}
    \caption{Depdive workflow}
    \label{fig:depdive}
\end{figure*}

In this section, we describe the implementation of the update audit tool, Depdive. Figure \ref{fig:depdive} shows a high-level workflow of the tool.

Depdive takes four arguments as input: 
\begin{enumerate*}
    \item registry name;
    \item package name;
    \item current version; and
    \item update version.
\end{enumerate*}
Depdive works with four package registries, namely Crates.io, npm, PyPI, and RubyGems. The input can be an update from any version to another version available in the registry. We scope Depdive's implementation to GitHub repositories, as we rely on data available through the GitHub platform to determine if a commit was code-reviewed, as explained in Section \ref{sec:reviewcheck}.

Depdive collects package code both from the registry and the repository and compares them to identify the phantom artifacts. Phantom artifacts in the registry do not map to any corresponding artifact in the repository, and therefore, cannot be audited for code review without a provenance tracking mechanism. Hence, Depdive outputs phantom artifacts separately from the code review analysis. Algorithm \ref{alg:step1} lists pseudocode for determining phantom files and lines in an update.

Afterward, Depdive maps each line of code changed in the update to its corresponding commit in the repository that made the change. Algorithm \ref{alg:step2} lists the pseudocode for identifying the code changes between two versions and mapping them to their corresponding commits. Depdive then determines if the commit was code-reviewed through four GitHub-based checks. In the following subsections, we explain each step in detail.

\subsection{Collect package code}

In this subsection, we explain how Depdive collects package code from both the registry and the repository.

\subsubsection{Download package code from the registry:} Depdive downloads package code for both the current and the update version from the respective registries. While Crates.io, npm, and RubyGems provide every package in a uniform format, PyPI packages can be available in multiple formats in the registry. When multiple formats are available, we prefer the default \textit{wheel} distribution.

\subsubsection{Locate package repository:} 
We determine a package's repository from the metadata provided by the registry. Afterward, we locate the directory path of the package within the repository. 

A repository can contain source code for multiple packages. Therefore, we need to know the directory path of a package for an accurate one-to-one mapping of files between the registry and the repository. For example, the filepath \textit{CHANGE\-LOG.md} in the Rust package \textit{tokio} maps to the filepath \textit{tokio/\-CHANGELOG.md} in the repository.

For Crates.io, npm, and RubyGems, Depdive identifies the directory path by locating the manifest file (\textit{Cargo.toml}, \textit{package.json}, and \textit{Gemspec}) of the package in the repository. 
While PyPI packages do not contain a uniform manifest file, we locate the directory by matching the filepaths in the registry with the filepaths in the repository. Through directory path locating, Depdive also validates the retrieved repository for a package. Further, a repository can contain submodules that point to different repositories. We obtain commit history recursively for all the submodules in the repository.

\subsubsection{Map file path from the registry to the repository:} 
For each file in the registry, we obtain the repository filepath by combining the package directory and the registry filepath. We follow this simple heuristic of filepath matching based on the assumption that packages are bundled by maintaining the same directory structure in the repository. We have found this assumption to be generally true in our study. Further, we also resolve the cases where a filepath contains a symbolic reference to another file within the repository.

\subsection{Identify release commit} 
\label{sec:releasecommit}
To compare the package code in the registry and the repository, we need to identify the head commit from which a certain version was built and uploaded in the registry. We refer to this commit as the \textit{\textbf{release commit}}. 

Depdive identifies the release commit through the git tags in the corresponding repository. Tagging the release commit with the version number is a recommended developmental practice~\cite{releasetag} and was followed in prior research work~\cite{goswami2020investigating, imtiaz2021open}. Depdive identifies the release tag and the associated commit for a version through a regular expression match. If the repository does not contain a single tag to match the given version, Depdive fails to analyze the corresponding update. 

While not every repository annotates the release commit via git tags or may do it inaccurately, the alternative is to compare the repository code with the registry code at all the commits in the history and take the commit with the smallest difference. Registry code may contain phantom artifacts which will not be present in the release commit, and therefore, may not return an exact match with any single commit. Vu et al. developed LastPyMile~\cite{vu2021lastpymile} to identify phantom artifacts in a PyPI package by comparing the registry code with every commit in the repository. But their approach cannot identify the specific release commit for a given version. Therefore, we take the heuristic to identify release commits via repository tags. 

% of the tag name with the version number and the package name. 
% In the case where the regex returns more than one match, we further check if the tag name contains the package name, as the repositories with multiple packages may have different releases for each package.

\subsection{Identify phantom artifacts}
Depdive compares the diff~\footnote{\url{https://git-scm.com/docs/git-diff}} between the package code of the current and update version in the registry, and the diff between the two versions' corresponding release commits in the repository. However, code changes can be present in the registry, but not in the repository. We refer to such changes as phantom artifacts~\cite{vu2021lastpymile}. Depdive outputs phantom artifacts in two categories, \textit{phantom files}, and \textit{phantom lines}:

\subsubsection{\textbf{Phantom file:}} We define phantom files in an update as files that are present in the update version in the registry, but not present in the corresponding  repository filepath at the release commit (line 8-14 in Algorithm \ref{alg:step1}).

\subsubsection{\textbf{Phantom line:}} We define phantom lines in an update as the lines of code changes that are present in the diff between the current and the update version, but not present in the diff between the corresponding release commits (line 15-24 in Algorithm \ref{alg:step2}). 
% Note that, inaccuracy in identifying release commit for a version can output false positives in phantom output from Depdive.

% As phantom files and lines are not part of the development history between the two release commits, we are unable to audit them for code review without a mechanism for provenance tracking for these phantom artifacts.  Therefore, Depdive filters out phantom files and lines at this step, and outputs them separately from the code review analysis. 

\begin{algorithm}
\floatname{algorithm}{Step}
\caption{Identify Phantom artifacts in an update}
\label{alg:step1}
\begin{algorithmic}[1]
\REQUIRE registry name: $E$
\REQUIRE package name: $P$
\REQUIRE current version: $X$
\REQUIRE update version: $Y$

\STATE $P_X = DownloadRegistryCode(E, P, X)$
\STATE $P_Y =  DownloadRegistryCode(E, P, Y)$
\STATE $F_{YP} = GetPackageFilepaths(P_Y)$

\STATE $R, D = LocateRepositoryAndDirectory(E,P)$
\STATE $C_X = IdentifyReleaseCommit(R, X)$
\STATE $C_Y = IdentifyReleaseCommit(R, Y)$
\STATE $F_{YR} = GetPackageFilepathsAtCommit(R, D, C_Y)$

\STATE Set of phantom files, $H_{PF} = \emptyset $
\FORALL{$ f \in F_{YP}$} 
\STATE $f_r = GetRepositoryFilePath(f, D)$
\IF{$f_r \not\in F_{YR}$}
\STATE $H_{PF}  = H_{PF}  \cup f $
\ENDIF
\ENDFOR

\STATE Map of phantom lines to files: $M_{PL}$
\FORALL{$ f \in F_{YP} - H_{PF}$} 
\STATE $f_r = GetRepositoryFilePath(f, D)$
\STATE Code changes in the registry, 
\STATE $D_P = Diff(f, P_X, P_Y)$
\STATE Code changes in the repository, 
\STATE $D_R = Diff(f_r, R, C_X, C_Y)$
\STATE phantom lines, $P_f = D_P - D_R$
\IF{$P_f \neq \emptyset$}
\STATE $M_{PL}.insert(f, P_f)$
\ENDIF
\ENDFOR
\RETURN Phantom files and phantom lines, $H_{PF}, M_{PL}$
\end{algorithmic}
\end{algorithm}

\begin{algorithm}
\floatname{algorithm}{Step}
\caption{Map code delta to corresponding commits in the repository}
\label{alg:step2}
\begin{algorithmic}[1]
%\REQUIRE Repository: $R$
\REQUIRE Release commit of current version: $C_X$
\REQUIRE Release commit of update version: $C_Y$
\REQUIRE Files changed in update: $F_{XY}$

\STATE Commits between $C_X$ and $C_Y$, 
\STATE $L_{XY} = git\_log (C_X..C_Y)$
\STATE Common ancestor,  $C_A = ParentOf(Oldest(L_{XY})$
\STATE Map of added lines to commit: $A_{XY}$
\STATE Map of removed lines to commit: $R_{XY}$
\FORALL{$f \in F_{XY}$}
\STATE Git blame: $B_f^Y = git\_blame(f, C_Y)$
\FORALL{$(l,c) \in B_f^Y$}
\IF{$c \in L_{XY}$}
\STATE $A_{XY}.insert(f,(l,c))$
\ENDIF
\ENDFOR
\STATE Reverse git blame: 
\STATE $RB_f^X = git\_blame(f, C_A, reverse=True)$
\FORALL{$(l,c) \in RB_f^X$}
\IF{$c \neq C_Y$}
\STATE $rc = findRemovalCommit(f,c,l)$
\IF{$rc \in L_{XY}$}
\STATE $R_{XY}.insert(f,(l,rc))$
\ENDIF
\ENDIF
\ENDFOR

\ENDFOR
\RETURN Code changes to commit map, $A_{XY}, R_{XY}$
\end{algorithmic}
\end{algorithm}

\newcommand\samebranch{%
\begin{tikzpicture}
\coordinate (a) at (0,0);
 \coordinate (b) at (2,0);
% nodes
\node [draw, circle] at (a) (A) {$C_X$};
\node [draw, circle] at (b) (B) {$C_Y$};
% arrows
\draw [->] (A) -- (B);
\end{tikzpicture}%
}

\newcommand\diffbranch{%
\begin{tikzpicture}
\coordinate (a) at (0,0);
 \coordinate (b) at (2,.5);
  \coordinate (c) at (2,-.5);
% nodes
\node [draw, circle] at (a) (A) {$C_A$};
\node [draw, circle] at (b) (B) {$C_X$};
\node [draw, circle] at (c) (C) {$C_Y$};
% arrows
\draw [->] (A) -- (B);
\draw [->] (A) -- (C);
\end{tikzpicture}%
}

\begin{figure}
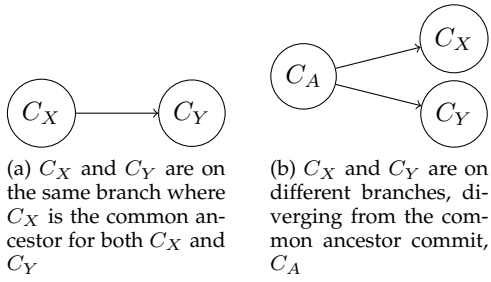

\centering

\subfloat[$C_X$ and $C_Y$ are on the same branch where $C_X$ is the common ancestor for both $C_X$ and $C_Y$]{\label{fig:commit_a}\samebranch}
%\label{fig:commit_a}
\hspace{0.5cm}
%\par
\subfloat[$C_X$ and $C_Y$ are on different branches, diverging from the common ancestor commit, $C_A$]{\label{fig:commit_b}\diffbranch}

\caption{The relative position of the release commit of version X ($C_X$) and the release commit of version Y ($C_Y$) in the repository commit history.}
\label{fig:commit}
\end{figure}

\subsection{Map code delta to commits}
\label{sec:codedelta}
Besides phantom files and lines, the rest of the diff between two versions in the registry maps to the diff between the corresponding release commits in the repository. However, not every line of change in the diff can be mapped to a specific commit within the two release commits. To explain why, consider the two commit history graphs shown in Figure \ref{fig:commit}, where $C_X$ and $C_Y$ refer to the release commit for version X and Y, respectively. In Figure \ref{fig:commit_a}, both $C_X$ and $C_Y$ lie on the same development branch, where $C_X$ is a direct ancestor of $C_Y$. In this case, each line of change in the diff between $C_X$ and $C_Y$ can be mapped to a single commit between those two commits. Contrarily, in Figure \ref{fig:commit_b}, $C_X$ and $C_Y$ lie on two different branches with a common ancestor at commit $C_A$. In this case, a line of code that was added in $C_X$, but was not present in $C_A$ and subsequently in $C_Y$, will be shown as a deleted line in the diff between $C_X$ and $C_Y$ (and between the two versions fetched from the registry). However, that line of code was not deleted in any commit, rather just did not appear between $C_A$ and $C_Y$.

Therefore, we define \textit{\textbf{code delta}} in an update as code changes that can be mapped to a commit that is present in the update version, but not present in the current version. 
Following this definition, we only consider the commits between $C_A$ and $C_Y$ in Figure \ref{fig:commit_b} as the commits between $C_X$ and $C_Y$ (double-dot commit range as per git syntax). The rationale behind this approach is that when analyzing an update, Depdive assumes that the current version is already trusted by the user, and therefore, focuses analysis only on the code changes in the new update.
Consequently, in Figure \ref{fig:commit_b}, we assume that if  $C_X$ is trusted by a user, any ancestor commit of $C_X$ is also trusted by the user. Therefore, when updating from $C_X$ to $C_Y$, we only focus on the code changes between their common ancestor $C_A$ and the update version $C_Y$. In the simpler case, where both $C_X$ and $C_Y$ lie on the same branch as shown in Figure \ref{fig:commit_a}, $C_X$ itself is the common ancestor of the two versions.

Algorithm \ref{alg:step2} shows Depdive's implementation of identifying the code delta between two versions and mapping each line of code in the delta to a corresponding commit. To identify the newly-added lines in the update version, Depdive runs \textit{git blame} on each package file at commit $C_Y$ to obtain the commits that added each line in the file. If a line was added in a commit that lies between $C_X$ and $C_Y$, we take the line as a newly-added line in the update (line 6-10 in Algorithm \ref{alg:step2}). To identify the lines that have been removed in the update, we run \textit{reverse git blame} on each package file at commit $C_A$ to obtain the last commit where each line in the file was present. If a line was still present in $C_Y$, the line was not removed in the update version. For the rest of the removed lines, we identify the removal commit by traversing forward from the blamed commit up to $C_Y$. If the removal commit lies between $C_X$ and $C_Y$, we take the line as a removed line in the update (line 12-20 in Algorithm \ref{alg:step2}).
~\footnote{In the case where the two release commits lie on the same branch (Figure \ref{fig:commit_a}), the code delta in the update will be equal to or greater than the code diff between the versions fetched from the registry. The code delta can be greater in cases where a line was modified after the release commit of the current version, but was reverted before the update version. Such a line will not appear in the code diff between versions from the registry, but will appear in the code delta calculated by Depdive.}

Along the process, we also obtain the commit mapping for each newly added and removed line of code in the update. Depdive also handles corner cases where a file has been renamed; a file that already existed in the repository but was newly included in the registry; and a file within a submodule pointing to a different remote repository. 

\subsection{Determine if a commit was code-reviewed}
\label{sec:reviewcheck}
After obtaining the commits involved in an update, we need to determine if the commit was code-reviewed. Git does not store data on code-review for a commit. Therefore, we focus on the projects that use GitHub as their development platform and look for evidence on GitHub to determine if a commit had been code-reviewed. Specifically, we apply four checks. Note that, we adopt these checks from ``Security Scorecards''~\cite{ossfscorecard}, a tool that analyzes the last 30 commits of a GitHub project and generates a score for a project's adherence to code review. Below, we explain the four checks:

\begin{enumerate}
    \item \textit{GitHub review}: We check if a commit belongs to a pull request and if the pull request was reviewed through GitHub's native code review tooling.
    \item \textit{Different merger}: We check if a commit belongs to a pull request and if the pull request was opened and merged by two different GitHub accounts. 
    % Contributors, who do not have direct write access to a project, open pull requests on GitHub to submit their code changes, and maintainers can merge these pull requests without providing explicit approval through GitHub's code review tooling. However, we assume that the pull request was reviewed by the merger before acceptance in such cases.
    \item \textit{Different committer}: We check if a commit was authored and committed by two different accounts. 
    \item \textit{Third-party tools}: Besides GitHub's native code review tooling, a project can use external code review tools, e.g., Gerrit~\cite{gerrit}, or custom bots to handle code-review, e.g., Prow~\cite{Prow}. When a commit is reviewed on Gerrit, the commit message contains metadata for the Gerrit review. When a pull request on GitHub is handled by Prow, Prow adds a label to the pull request to indicate if it was code-reviewed. Following Security Scorecards' implementation, we also check for evidence for Gerrit or Prow review in Depdive. 
\end{enumerate}

We determine a commit as code-reviewed if any one of the four checks is met.
When checking for different mergers or committers, we do not consider a commit to be code-reviewed if both the author and reviewer are bot accounts on GitHub. We further exclude cases where GitHub's own bots are involved (e.g., GitHub actions). However, in cases where a pull request has been opened by a bot, e.g., Dependabot~\cite{dependabot}, and merged by a non-bot account, we consider the commit as code-reviewed. 
% Similarly, bots are often used to handle pull-request acceptance after checking for certain criteria, such as code-review. If a pull request has been opened by a non-bot account and merged by a bot account, e.g., bors~\cite{bors}, we consider the commit as reviewed. 

\begin{table*}
    \centering
    \begin{tabular}{lrrrr}
        Registry & \makecell{No. of \\Selected Packages} & \makecell{No. of \\Selected Updates} & \makecell{No. of Packages \\analyzed by Depdive} & \makecell{No. of Updates \\analyzed by Depdive}   \\
\hline
 Crates.io  &  990 &  8,434 & 833 (84.1\%)  & 6,326 (75.0\%)  \\
 npm        &  989 &  8,158 & 919 (92.9\%)  & 7,178 (88.0\%)  \\
 PyPI       &  992 &  8,657 & 788 (79.4\%)  & 6,089 (70.3\%)  \\
 RubyGems   &  991 &  8,646 & 674 (68.0\%)  & 5,351 (61.9\%)  \\
 \hline
 Total      & 3,962 & 33,895 & 3,214 (81.1\%) & 24,944 (73.6\%) \\
\hline
    \end{tabular}
    \caption{Updates successfully analyzed by Depdive from the selected packages}
    \label{tab:dataset}
\end{table*}

\section{Dataset}
\label{sec:dataset}
In this section, we explain the packages and their updates we selected for this study, and the updates that Depdive could successfully analyze  based on which we complete our empirical analysis.

\subsection{Package selection}

We select the latest ten updates of the most downloaded 1000 packages in each of the four package registries, namely Crates.io, npm, PyPI, and RubyGems.
For Crates.io and RubyGems, we download the official data dumps that are updated daily and then select the top packages in order of the total download count across all versions. For PyPI, we use the dataset from \cite{hugo_van_kemenade_2022_5812615} that is updated on a monthly basis. We collected the data for these three registries at the end of December 2021. For NPM, we use the dataset from ~\cite{zahan2021weak} that was constructed in August 2021. While the download count can be inflated in different ways, including through CI/CD tooling, sampling the most downloaded packages to study a package ecosystem is an established approach~\cite{vu2021lastpymile, bommarito2019empirical} and provides an estimation of the most used packages in a registry.

We then collect the list of the available version releases on the registry for each of the selected packages at the end of December 2021. Depdive can run on an update from any version to another (e.g., 1.8.3 → 1.8.4, 1.5.1 → 1.9.0, 1.0.0 → 3.0.3 etc.). However, for this empirical study, for each version, we consider an update only from its prior version according to SemVer~\cite{semver} ordering. For example, for \textit{tokio@1.8.4}, we consider an update from \textit{tokio@1.8.3} to \textit{tokio@1.8.4} to pass as input to Depdive. We choose this approach to avoid any data duplication, that is, the same code changes and commits to appear in multiple updates of a package selected in this study. Further, we exclude any pre-release version from our data set.

We also restrict our data set to only the most recent ten updates for each of the packages. The rationale is that Depdive makes multiple GitHub API calls for each commit in an update for code-review checks. However, GitHub limits API calls to 5000/hr for each user, which puts a constraint on how much analysis we can do within an hour. However, choosing at least ten updates will provide us with the data to measure the consistency of a package in its code review coverage in each update. In the case where a package has less than ten updates, we consider all of them in our study.

Out of the 4000 packages, 26 packages had only one release listed on the registry. Further, 12 packages had only one regular release available, while the other available versions were pre-releases. We exclude these 38 packages as there were no updates available to analyze. Finally, we select 33,895 updates from 3,962 packages. Table \ref{tab:dataset} shows a breakdown of these initially selected updates across the four registries. 

\subsection{Depdive analysis}
\label{sec:metrics}
Out of the selected 33,895 updates, Depdive successfully analyzed 24,944 (73.6\%) updates. In Section \ref{sec:empirical}, we present our empirical analysis based on Depdive's output for these 24,944 updates. Table \ref{tab:dataset} shows a breakdown of the dataset of this study. We collect the following metrics for each update to conduct our empirical analysis:
\begin{enumerate*}
    \item No. of phantom files,
    \item No. of files with phantom lines,
    \item No. of added phantom lines~\footnote{While Depdive also outputs phantom removal, we observe that the removed lines are often also phantom lines from the old version.}, and
    \item Code Review Coverage (\textit{CRC}).
\end{enumerate*}

Note that, we measure the \textit{CRC} of an update as the proportion of the update's code delta that has been code-reviewed. Here, the code delta, as explained in Section \ref{sec:codedelta}, is the code changes in an update that can be mapped to a commit in the repository and therefore, can be classified if code-reviewed or not.

% \begin{itemize}
%     \item \textbf{No. of phantom files:} The number of files in the update version present in the registry, but not in the repository.
%     \item \textbf{No. of files with phantom lines:} The number of files containing changes in an update that are not present in the repository.
%     \item \textbf{No. of added phantom lines:} The number of phantom lines newly added in the update version.
%     \item\textbf{Code Review Coverage (\textit{CRC}):} As explained in Section \ref{sec:codedelta}, Depdive identifies code delta in an update as lines of code (LOC) changes between the current and the update version that can be mapped to a commit in the repository and therefore, can be classified if code-reviewed or not. We measure code review coverage 
%     % \item \textbf{Commit review coverage:} Depdive also outputs the number of commits and their unique identifiers that made the code changes in an update. We measure commit review coverage as the proportion of the commits in an update that has been code reviewed. 
%     % \item \textbf{Code Review Category:} For each code-review commit, we also categorize the code review type based on the checks listed in Section \ref{sec:reviewcheck}.
% \end{itemize}
% In Section \ref{sec:empirical}, we report our empirical analysis based on the metrics listed above.
\subsection{Why Depdive failed}
As shown in Table \ref{tab:dataset}, Depdive only successfully analyzed 73.6\% of the initially-selected updates. 
% Table \ref{tab:failure} shows a breakdown of the reasons behind Depdive's failure for the rest.
The primary reason that Depdive could not analyze an update was the absence of git tags in the repository pointing to the release commit of a version, which was the case for 5,747 updates (17.0\%). The next major reason is Depdive's inability to either locate or validate the repository of a package, which was the case for 2,669 of all the updates (7.9\%). For 127 updates (0.4\%), the listed repository was not on GitHub. For the rest of 1.1\% of the updates, Depdive failed for various reasons, including not being able to read a file containing non-Unicode characters, a private submodule within the repository, and version code not being available on the registry.

If Depdive were to be deployed in a CI/CD pipeline, our empirical evaluation shows that the tool could fail in 26.4\% of the cases. However, the failure rate may be reduced if the package repositories followed the best developmental practices, such as annotating a git tag for each released version.
We recommend that package manager tools add a feature to include metadata on the repository, package directory, and release commit in the package bundle during the build process to aid in a third-party audit.

\section{Findings}
\label{sec:empirical}

Based on Depdive's output, we answer the following two research questions for the four studied registries: \textit{\textbf{RQ1:}} To what extent do phantom artifacts exist in the updates of the most downloaded packages?  
and \textit{\textbf{RQ2:}} Excluding phantom artifacts, what is the code review coverage in the updates of the most downloaded packages?
In the following two subsections, we present our findings:

\subsection{RQ1: Phantom artifacts}
\label{sec:phantomres}

\begin{table*}[]
    \centering
    \begin{tabular}{lrrrrr}
        Registry &
         \makecell{Total\\ packages} & \makecell{Total\\ Updates} & \makecell{ Packages with\\ phantom files} & \makecell{ Updates  with\\ phantom files} & \makecell{ Median phantom\\ file count} \\
\hline
 Crates.io  &  833 &  6,326 & 42 (5.0\%)     & 123 (1.9\%)   &            3.25 \\
 npm        &  919 &  7,178 & 413 (44.9\%)   & 2,489 (34.7\%) &            2.00    \\
 PyPI       &  788 &  6,089 & 355 (45.1\%)   & 2,088 (34.3\%) &            1.50  \\
 RubyGems   &  674 &  5,351 & 71 (10.5\%)    & 323 (6.0\%)   &            2.00    \\
 \hline 
 Total      & 3214 & 24944 & 881 (27.4\%)   & 5023 (20.1\%) &            2.00    \\
\hline
    \end{tabular}
    \caption{Updates with phantom files, i.e., files present in the registry but not in the repository.}
    \label{tab:phantomfile}
\end{table*}
\begin{table*}[]
    \centering
    \begin{tabular}{lrrrrrrr} Registry &
        \makecell{Total\\packages} & \makecell{Total\\Updates} &   
          \makecell{ Packages with\\ phantom lines} & \makecell{ Updates  with \\phantom lines} & \makecell{Median file count\\ with phantom lines} &
         \makecell{Median count of \\added phantom lines} \\
\hline
  Crates.io  &  833 &  6,326 & 63 (7.6\%)   & 87 (1.4\%)    &                          1.0 &               2.0 \\
 npm        &  919 &  7,178 & 265 (28.8\%) & 1,733 (24.1\%) &                          1.0 &               1.0 \\
 PyPI       &  788 &  6,089 & 157 (19.9\%) & 471 (7.7\%)   &                          1.0 &               3.0 \\
 RubyGems   &  674 &  5,351 & 105 (15.6\%) & 185 (3.5\%)   &                          1.0 &               2.0 \\
 \hline
 Total      & 3,214 & 24,944 & 590 (18.4\%) & 2,476 (9.9\%)  &                          1.0 &               2.0 \\
\hline
    \end{tabular}
    \caption{Updates with phantom lines, i.e., changes in a file in the registry but not in the repository}
    \label{tab:phantomline}
\end{table*}
Table \ref{tab:phantomfile} shows the occurrence of phantom files, and Table \ref{tab:phantomline} shows the occurrence of phantom lines in updates across the four studied registries. Overall, 24.9\% of the updates contained either a phantom file or a phantom line (3.2\%, 44.3\%, 38.5\%, and 9.1\% in the case of Crates.io, npm, PyPI, and RubyGems updates, respectively). Below, we discuss our findings:

\subsubsection{\textbf{Phantom files:}}
We find that 20.1\% of the analyzed updates and 27.4\% of the analyzed packages had at least one phantom file. We find that npm and PyPI updates are more likely to contain a phantom file (34.7\% and 34.3\%, respectively) than Crates.io and Ruby Gems updates (1.9\% and 6.0\%, respectively). Overall, we identified 306,940 phantom files across all the updates. Below, we provide a crude characterization of the phantom files that we identified:

Across all the npm updates, we identified 123,000 phantom files. We observed that for some npm packages, the code in the registry could be a transpiled version of the source code in the repository. For example, the source code may be written in TypeScript, whereas the package contains transpiled JavaScript code. Similarly, npm packages can also contain minified JavaScript, or JavaScript transpiled through transcompilers such as \textit{babel}. Overall, 89.4\% of the phantom files in the npm updates are JavaScript files (\textit{.js, .d.ts, .cjs, .mjs, .min.js, .map, .flow} files) which may be code transpiled during the package build process. Besides transpiled code, 9.5\% of the npm phantom files are \textit{.json} files which are either data or configuration files, and possibly were git-ignored in the repository through the \textit{.gitignore} file.

We identified 146,564 phantom files across all the PyPI updates. The majority of these files (46.1\%) are machine-generated header files for C/C++ (\textit{.h, .hpp, .inc} files), while 7.4\% of the phantom files are compiled binaries (\textit{.so, .jar, .dylib}). The core engine of many Python packages, e.g., \textit{tensorflow}, are written in languages like C/C++, while the Python files only provide an interface to communicate with the engine. While the repositories of these packages contain the source code for the engines, the Python packages, in their default \textit{wheel} distribution, only contain the compiled binaries and the machine-generated header files alongside the Python source files. The header files and binaries come from 118 distinct packages in our dataset (15.0\% of the analyzed PyPI packages).
Moreover, we find 16.3\% of the phantom files to be Python files, where one-third of them (32.1\%) are \textit{\_\_init\_\_.py} files which are presumably machine-generated. Finally, we have also observed non-Python code files (7.1\% are \textit{.js, .ts} files), data files (5.0\% are \textit{.dat} files), and git-ignored files (e.g., \textit{.pyc, .pyi, .py~} files) as phantom files.

Across all the RubyGems updates, we found 7,483 phantom files. However, 83.2\% of these files came from 7 packages where the package either (i) bundled its own Ruby dependency packages, (ii) bundled non-Ruby dependency code, or (iii) included log files that were git-ignored in the repository. 

Across Crates.io updates, we identified 29,893 phantom files. However, 84.4\% of these files come from a single package that contained the source code for its project website hosted in a repository branch. Overall, across all four registries, we have observed files that are git-ignored in the repository to be included in the registry, e.g., \textit{.DS\_Store, .npmignore, .editorconfig, setup.cfg} files.

\vspace{1em}
\textbf{Accidental vs legitimate phantom files:}
Broadly, we can characterize the phantom files into two categories: (i) phantom files due to legitimate reasons such as compiled binaries in PyPI \textit{wheel} distribution, transpiled JavaScript, auto-generated files, and third-party dependency code; and (ii) phantom files presumably put in mistakenly, such as the git-ignored files.

For the legitimate phantom files, we need to track their provenance, and then determine if changes in the origin files have been code-reviewed or not. We discuss current research efforts and our recommendations on handling legitimate phantom files in Section \ref{sec:discussion}. On the contrary, we recommend package maintainers issue a new clean release in case of accidental phantom files, as the \textit{last mile} between the repository and the registry has been used as an attack vector in the past to sneak in malicious code~\cite{vu2021lastpymile, ohm2020backstabber}. 

Overall, while there may be legitimate reasons behind phantom files, we cannot audit the changes in these files through our proposed approach. Package users may manually review these files before accepting an update, especially when the current version does not contain any phantom files, but the new update does. In our dataset, 461 packages (52.1\% of all the packages with phantom files) had phantom files only in a subset of all their updates. We presume that these packages may have accidentally put in phantom files in some of their updates. 

\begin{table*}[]
    \centering
    \begin{tabular}{lrrrrrr}
        Registry & \makecell{Total\\Package} & \makecell{Total\\Updates} & \makecell{Median LOC changes\\in updates} & \makecell{Median code review\\ coverage (\textit{CRC}) in updates}
        %  & \makecell{Median commit\\ review coverage\\ in updates} 
        &
         \makecell{Updates with \\ 100\% \textit{CRC}}&
         \makecell{Updates with \\ 0\% \textit{CRC}}\\
\hline
 Crates.io  &  830 &  6,293 &       75.75 & 42.9\% & 915 (14.5\%)  & 2,068 (32.9\%) \\
 npm        &  918 &  7,147 &       25.00    & 5.9\%  & 87 (1.2\%)    & 3,497 (48.9\%) \\
 PyPI       &  780 &  5,861 &       76.75 & 52.7\% & 1,425 (24.3\%) & 1,760 (30.0\%) \\
 RubyGems   &  672 &  5,309 &       61.25 & 31.0\% & 269 (5.1\%)   & 1,663 (31.3\%) \\
 \hline
 Total      & 3,200 & 24,610 &       51.00    & 27.2\% & 2,696 (11.0\%) & 8,988 (36.5\%) \\
\hline

    \end{tabular}
    \caption{Code review coverage of the analyzed updates
    }
    \label{tab:crc}
\end{table*}

 \begin{table}[]
     \centering
     \begin{tabular}{lrrr}
         Registry & \makecell{Total\\ Packages} & \makecell{Packages with \\
         all updates\\
         having 100\% \textit{CRC}} &
         \makecell{Packages with \\
         all updates\\
         having 0\% \textit{CRC}}\\
\hline
 Crates.io  &  830 & 99 (11.9\%)  & 217 (26.1\%) \\
 npm        &  918 & 5 (0.5\%)    & 386 (42.0\%) \\
 PyPI       &  780 & 162 (20.8\%) & 185 (23.7\%) \\
 RubyGems   &  672 & 23 (3.4\%)   & 161 (24.0\%) \\
 \hline
 Total      & 3,200 & 289 (9.0\%)  & 949 (29.7\%) \\
\hline
     \end{tabular}
     \caption{Packages with all updates fully code-reviewed, or not code-reviewed at all.}
     \label{tab:fullzero}
 \end{table}
\begin{figure}
    \centering
    \includegraphics[scale =0.5]{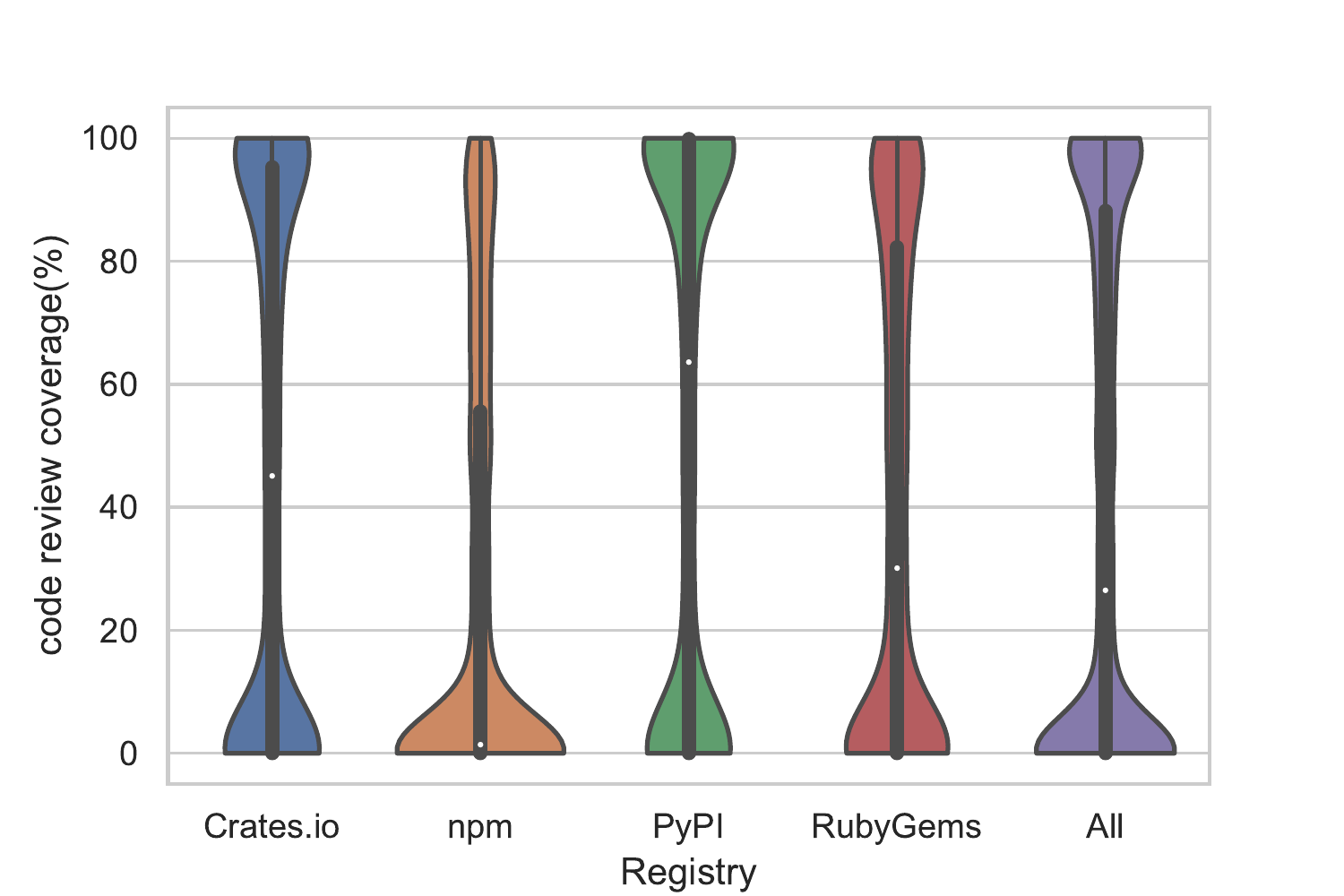}
    \caption{Violin plot of code review coverage across updates}
    \label{fig:violin}
    
\end{figure}

\subsubsection{\textbf{Phantom lines:}} We find that 9.9\% of the analyzed updates across 18.4\% of the packages had code changes that could not be mapped back to a commit in the repository. We find that phantom lines are generally small changes, with 2 added lines in 1 file at the median. We also find that npm updates are more likely to contain phantom lines than updates in the other three studied registries.

We identified phantom lines in 1,733 (24.1\%) npm updates, where 93.9\% of these updates contained phantom lines in the manifest file, \textit{package.json}, that was presumably dynamically generated with added data such as the release commit.
Further, as explained when discussing phantom files, the npm package code in the registry can be transpiled from the code in the repository. Therefore, the same filepath may have different code in the registry and the repository, resulting in many updates with phantom lines.

Across PyPI packages, we identified 471 updates (7.7\%) with phantom lines, where 51.6\% of the updates had phantom lines in a file named \textit{\_version.py} that was presumably dynamically generated with added data such as the release commit, the build date, and the version number. Further, \textit{\_\_init\_\_.py} files can also be generated dynamically during the build process and may differ in content from the repository copy, which resulted in phantom lines in 117 updates (24.8\%). For both Crates.io and RubyGems, we do not find any common pattern in the identified phantom lines. However, false positives may appear in the case where a release commit was inaccurately tagged. 

While some files can be dynamically generated during the build process, we find that phantom lines are less likely to occur due to legitimate reasons in Crates.io, PyPI, and RubyGems updates, and package users should manually review these lines before accepting an update. 

\begin{tcolorbox}
We find that phantom artifacts are not uncommon in package updates, specifically in the case of npm and PyPI packages (20.1\% of the updates having at least one phantom file). While further research is required to verify if the changes in phantom artifacts were code reviewed or not, we recommend practitioners remain careful of accidental phantom artifacts, such as publishing local files in the package that are not checked into the repository. 
\end{tcolorbox}

\subsection{RQ2: Code review coverage}
\label{crcfinding}

In this section, we present our findings on code review coverage (\textit{CRC}) for the analyzed updates. Note that, we exclude phantom artifacts in \textit{CRC} measurement, and only consider the changes in the package code that can be mapped to the repository. Table \ref{tab:crc} presents the 
median lines of code (LOC) changes and the median \textit{CRC} across updates in the four registries. We excluded 334 updates from the 24,944 analyzed updates as they contained zero code delta as measured by Depdive. These updates either only made changes in the non-package files in the repository, or only made changes in the phantom files. Further, we identified 110,657 commits as code-reviewed, of which 60.1\% were GitHub review, 31.5\% were Different merger, and 8.4\% were Different committer, as per the checks explained in Section \ref{sec:reviewcheck}.

The table also shows the portion of the updates with 100\% and 0\% \textit{CRC} across the four registries. We find that 11.0\% of the updates were fully code-reviewed, while 36.5\% of the updates were not code-reviewed at all. The rest of the 52.5\% of the updates were only partially code-reviewed. 
Further, we find that the median \textit{CRC} across all the analyzed updates stands at 27.2\%.

We find the npm packages to have the lowest median \textit{CRC}. In Section \ref{sec:phantomres}, we explained that npm packages may contain transpiled JavaScript, resulting in many phantom artifacts. In these cases, Depdive's  \textit{CRC} measurement would be limited to only a subset of the package files. For example, for the \textit{no-case} package, Depdive could only audit the package manifest files for code review. To address this limitation, we looked at the \textit{CRC} for the 3,947 npm updates that did not contain any phantom artifacts with a presumption that such updates are less likely to contain transpiled code. However, we still observe a low median \textit{CRC} of 6.1\% for these updates. While many npm packages are small and may only have a single maintainer, our analysis in this study may be an under-approximation for npm. The packages that contain transpiled JavaScript, e.g., packages with source code in TypeScript, may also be more likely to follow developmental best practices such as code review. We have observed npm packages from reputed  organizations like \textit{babel} and \textit{facebook} to develop their source code in TypeScript and adhere to code review in the majority of the commits. However, Depdive could only audit a subset of the files in these packages, and therefore, we may have got an under-approximated \textit{CRC} measurement. 

For Crates.io, PyPI, and RubyGems packages, we find the median \textit{CRC} to be 42.9\%, 52.7\%, and 31.0\%, respectively. Figure \ref{fig:violin} shows a violin chart of the \textit{CRC} across the updates. We observe an \textit{hourglass} shape in the violin chart, suggesting that updates tend to have either very high \textit{CRC} or very low \textit{CRC}. Open source packages are often maintained by a small group of maintainers. While the contributions from an outsider get code-reviewed by the maintainers, the code changes from the maintainers themselves may remain non-reviewed, therefore resulting in low-to-medium \textit{CRC}. On the contrary, we have found updates that can have very high \textit{CRC} yet not fully code-reviewed. For example, we have found 2,442 updates (9.9\%) that are larger than 51 LOC changes (overall median) and have greater than 90\% but less than 100\% \textit{CRC}.  

While most of the commits in these updates were code-reviewed, we found some commits did not go through the code review process, possibly due to the following two primary reasons as per our observation: (i) \textit{Non-critical changes:} the commit only changed configuration or documentation files; (ii) \textit{Cherry-picked commits:} the commit cherry-picked a commit from a different branch where it was code-reviewed either to backport a fix or to import changes to a release branch from the master branch. Note that, SLSA requires context-specific approval during a code review, which means the cherry-picked commits require their own separate review. Nonetheless, this phenomenon shows that non-reviewed code changes may sneak in even in packages that attempt to adhere to code review. A few notable packages where we found high \textit{CRC} but not 100\% \textit{CRC} include \textit{numpy, ansible-core} in PyPI, \textit{nokogiri} in Ruby, \textit{openssl} in Crates.io, etc.

Table \ref{tab:fullzero} shows packages for whom all the analyzed updates in our data set were measured to have 100\% \textit{CRC} or all the updates were measured to have 0\% \textit{CRC}. We find that only 9.0\% of the packages in our dataset had all their updates fully code-reviewed, most of them coming from PyPI and Crates.io. We observed that packages from reputed organizations such as \textit{google, Azure, rust-lang, tokio-rs}, and \textit{aws} are likely to consistently have fully code-reviewed updates. On the contrary, 29.7\% of the packages had none of their updates code-reviewed at all. These packages are typically maintained by a small group of maintainers who do not review each other's code.

\begin{tcolorbox}
We find that 52.5\% of the analyzed updates were only partially code-reviewed, with an overall median code review coverage (\textit{CRC}) of 27.2\%. Further, only 9.3\% of the analyzed packages were measured for 100\% \textit{CRC} across all their updates. We observe that updates tend to have either low \textit{CRC} where maintainers only code-review outsider contributors' pull requests, or have high \textit{CRC} with only commits that modify configuration or documentation files not going through a code-review process.
\end{tcolorbox}

% \begin{table}[]
%     \centering
%     \begin{tabular}{lrr}
%         Code Review Type & Count & Rate  \\
%         \hline
%  GitHub Review       & 66412 & 59.6\% \\
%  Different Merger    & 34944 & 31.4\% \\
%  Different Committer &  9998 & 9.0\%  \\
% \hline
%     \end{tabular}
%     \caption{Code review category}
%     \label{tab:crcategory}
% \end{table}

\section{Maintainer Agreement Survey}
\label{survey}

We conducted a survey to evaluate whether the studied packages' maintainers agreed with the Depdive analysis. We emailed maintainers of the packages and provided an analysis report of a randomly-selected update (of that package) and asked
two Likert scale questions based on the analysis: \begin{enumerate*}
    \item \textit{Do you agree with our analysis?}
    \item \textit{How often do you require code review?}
\end{enumerate*}
We also gave options for the respondents to explain their answers.

We sent out the survey through email to the maintainers of 945 packages. We received 96 responses, with a response rate of 10.2\%. We did not survey packages where \begin{enumerate*}
    \item we could not collect the maintainers' valid email addresses;
    \item we already emailed the maintainer querying on a different package they also own; and
    \item the package had phantom artifacts resulting in incomplete analysis on our part. 
\end{enumerate*}

\begin{figure}
    \centering
    \begin{tikzpicture}
\pie[ radius=2,
color = {
        green,
        yellow,
        red,
    }]{49.0/Agree ,
    34.4/Somewhat Agree,
    16.7/Disagree }
\end{tikzpicture}
    \caption{Maintainers' agreement rate with Depdive analysis.}
    \label{fig:agreement}
\end{figure}
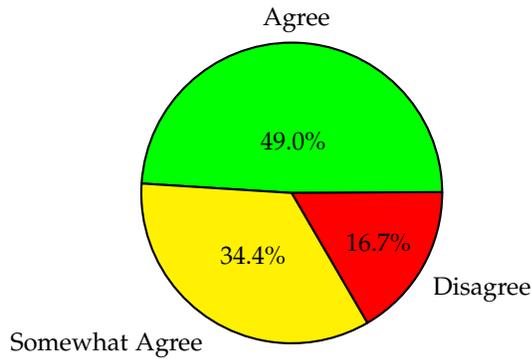

Figure \ref{fig:agreement} shows a pie chart of maintainer agreement to our analysis. 47 (49.0\%) respondents fully agreed to our analysis, while 33 (34.4\%) respondents partially agreed. The remaining 16 (16.7\%) respondents disagreed. In 48 cases, the maintainers provided the reasoning behind their disagreement. We classified the reasoning into the following categories: 

\begin{enumerate}
    \item \textbf{Non-functional changes (15):} Maintainers noted that the unreviewed commits did not change the source code of the package, rather only configuration files, e.g. package manifest files, CI/CD scripts, and documentation files, e.g. README.md, CHANGELOG.md. The maintainers disagreed with our analysis, noting that changes in these files should not be included in the \textit{CRC} analysis.
    
    \item \textbf{Single maintainer (8):} Maintainers disagreed on the rationale that \textit{CRC} analysis is invalid for their packages, as they are the sole maintainer. 
    
    \item \textbf{Review outside GitHub (7):} In 7 cases, maintainers noted that the commits flagged as unreviewed were indeed reviewed, either through an internal code review tool, or through discussion on GitHub.
    
    \item \textbf{Misunderstanding of the analysis (7):} There were 7 cases where maintainers disagreed for reasons that appear to be a miscommunication on our part regarding what the tool should do. For example, maintainers noted that Depdive missed some commits in the surveyed update. When investigating those updates, we found that the omitted commits were on non-package files, e.g., test files or CI/CD scripts, which were not present in the registry, and therefore, were not included in the \textit{CRC} analysis.  
    
    \item \textbf{Trivial changes (5):} Maintainers noted that the unreviewed commits only consisted of trivial changes, e.g., code re-formatting, and therefore, does not require a code review. They viewed flagging such commits as unreviewed as false alarms.
    
    \item \textbf{Others (6):} While we marked dependabot PRs merged by a developer as reviewed, two maintainers opined that they should be marked as unreviewed. In one case, the maintainer disagreed that the backported commits needed to be reviewed again. In another case, the unreviewed commit was only a merge commit. In one case, the package only included auto-generated content. In another case, the package is now unmaintained. 
\end{enumerate}

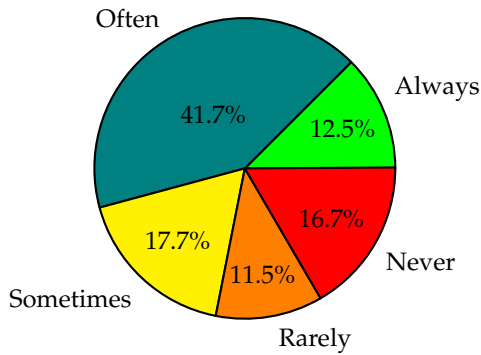
\begin{figure}
    \centering
    \begin{tikzpicture}
\pie[ radius=2,
color = {green,
        teal,
        yellow,
        orange,
        red,
    }]{12.5/Always,
    41.7/Often,
    17.7/Sometimes,
    11.5/Rarely,
    16.7/Never}
\end{tikzpicture}
    \caption{Maintainers' responses to code review requirement in their packages.}
    \label{fig:crrequirement}
\end{figure}

Finally, Figure \ref{fig:crrequirement} shows maintainers' responses to how often they require code review for new contributions in their package. We observe that the responses are in alignment with our empirical findings in Table \ref{tab:fullzero}. Only 12.5\% of the packages always require code review, while 41.7\% of the packages often require code review. On the other hand, 28.2\% of the packages either never or rarely require a code review. 

\begin{tcolorbox}
We found that maintainers largely agree with our analysis. While there are cases where the code review took place outside GitHub (and therefore, flagged as unreviewed by Depdive), the maintainers' disagreement primarily stemmed from their opinion that some changes (e.g., non-functional changes) should not be included in the \textit{CRC} analysis.
\end{tcolorbox}

\section{Limitations}
\label{limitation}
In this section, we discuss the limitations of our study.

\subsection{Reliability threat} We adopt multiple heuristics in designing Depdive that may result in false-positive outputs. Firstly, we map package files in the registry to a file in the repository through filepath matching. However, maintainers may choose a build process where the filepaths will be altered in the bundled package without altering the content of the file, in which case Depdive will output them as phantom files. 
% Similarly, we may fail to validate a correct repository and locate the package directory when the directory structure in the registry code and the repository do not match. 
While an alternate approach could be a pairwise comparison of all the files, our simpler heuristic is based on a realistic assumption that bundled packages typically follow the same directory structure as that in the repository. Secondly, we leverage repository tags to determine the release commit for a package version, inaccuracy in which may incorrectly output phantom artifacts. We explained this design decision in Section \ref{sec:releasecommit} and recommend packages themselves contain the metadata for the release commit.

Further, Depdive scopes its code-review checking to the GitHub platform, and therefore, may output false negatives in the case where a project is hosted on the GitHub platform but perform its code review elsewhere. Indeed, 7.9\% of the survey respondents listed that the code marked as unreviewed was actually reviewed in a way not covered by our checks. Overall, we chose our heuristics to keep the design simple while minimizing the possibility of false negatives.  In this regard, our empirical analysis of the phantom artifacts may represent an over-approximation, while the analysis of code review coverage may represent an under-approximation. 

\subsection{Generalizability threat} Our work studies the most downloaded packages in four package registries. However, as shown in Table \ref{sec:dataset}, we failed to analyze 26.4\% of the initially-selected updates. The failure may introduce unknown sampling biases in our dataset for the empirical analysis. Further, the threat to the \textit{CRC} analysis for the npm ecosystem due to the presence of transpiled code has been explained in Section \ref{crcfinding}. Nonetheless, we believe our empirical findings provide an evaluation of our proposed approach and offer many insights into the existence of phantom artifacts and \textit{CRC} among the most downloaded packages. 

\subsection{What should (not) be included in \textit{CRC}?}
In our survey, maintainers have opined that not all files should be included in the \textit{CRC} analysis. For example, multiple maintainers have opined that changes in the package manifest file, e.g., \textit{package.json} for npm, need not be reviewed. However, in the past, supply chain attacks were carried out by introducing a malicious package in the dependency chain listed in the manifest file~\footnote{https://security.snyk.io/vuln/SNYK-JS-NODEIPC-2426370}. 
% On the other hand, we have observed that many packages do not include configuration scripts and test files in the package uploaded in the registry. 
Similarly, changes in configuration scripts and test files were identified as \textit{non-code}, \textit{non-functional}, \textit{bookkeeping} changes by the survey respondents. Maintainers may follow the practice of not uploading these files in the registry if the files are not part of the package source code. Alternatively, one survey respondent suggested providing a feature for the maintainers to define an \textit{allowlist} that should be excluded from the \textit{CRC} analysis.

In our study, we have taken a conservative approach of including every change in the package code in the \textit{CRC} analysis. While not all changes may pose a security risk, future research is required to formally verify which files/changes are risk-free in each ecosystem, and can be safely ignored in any update audit.  

\section{Discussion}
\label{sec:discussion}
In this section, we discuss various aspects of Depdive's design and the implication of the findings from our empirical study:

\subsection{Depdive Design Philosophy}
One driving philosophy behind Depdive is that every line of code in the software supply chain should be traceable to at least two owners~\cite{slsa}. To achieve that, we need to map the package code downloaded from the registry to the commits in the repository, and then reliably determine the author and reviewers of the mapped commits. However, we face two challenges in the process:

\textbf{Provenance of the legitimate phantom artifacts:} We find that legitimate phantom artifacts, such as the programmatically generated files, can appear in a package. Especially, transpiled JavaScript code in the npm packages and compiled binaries in the PyPI packages make Depdive's audit incomplete without ensuring the integrity of the origin of these files.
Downstream projects can address this issue by retrieving package code directly from the source repository and then building the package by themselves. PyPI has a feature to provide a source distribution alongside the default \textit{wheel} distribution~\cite{pythondist} that contains the source code to compile the required binaries on the client's end. Clients then can run Depdive on the source distribution for a complete \textit{CRC} measurement. We recommend npm also consider this distribution model. 

Another way is to divide the update audit into two parts: (i) have a framework, such as \textit{in-toto}~\cite{torres2019toto} and \textit{Reproducible Builds}~\cite{reprobuild}, to verify that the package has indeed come from a certain commit point in the repository, and (ii) measure \textit{CRC} directly from the repository (without the step in Algorithm \ref{alg:step1}). However, these approaches have their own pros and cons. We recommend future research to investigate designing a package distribution model that minimizes the efforts required from the stakeholders while ensuring the integrity of the packages. 

\textbf{Quality of code review:} A rogue/hijacked maintainer account can easily bypass Depdive checks by creating a pull request from a sock account and then reviewing and accepting that pull request. SLSA requirement~\cite{slsareq} states that two trusted persons should be involved in a code review. Therefore, a tool like Depdive should also consider the digital identity of the involved developer accounts and flag any suspicious reviews. Similarly, the diversity and expertise of the reviewers should also indicate the overall quality of a code review. We consider providing a quality rating for each code review and flagging suspicious reviews as future work for Depdive.

\textbf{Difference with other tools:} There are existing tools that either (i) measure the code review adherence of a project;or (ii) identify phantom artifacts in a package. "Security Scorecard"~\cite{ossfscorecard} gives a score for a GitHub project's code review adherence by looking at the branch protection rules and the review history of the last 30 commits. While a periodic check can help in selecting a dependency, we aim to measure the code review coverage during \textit{each update}, which presents an added challenge of mapping the code changes from the registry to the repository commits. Further, Vu et al.~\cite{vu2021lastpymile} have developed LastPyMile to identify phantom artifacts in a PyPI package by comparing the package code with every commit in the repository. While Depdive's approach has methodological differences with LastPyMile, as explained in Section \ref{sec:releasecommit}, both the tools should return similar results (for phantom artifacts in PyPI updates). In a way, Depdive brings the above two tools' objectives into a single workflow and provides an isolated audit only for the changes in an update. Such an audit will help package users focus their review effort on the incremental changes in each update.

\subsection{The State of the Package Ecosystems}

Through our empirical evaluation of Depdive, we also present our findings on the code review coverage in the recent updates of the most downloaded packages. Below, we discuss some implications of our findings:

\textbf{The \textit{hourglass} phenomenon:} We have seen that packages either tend to have very high \textit{CRC} or very low \textit{CRC}, as depicted by an hourglass shape in Figure \ref{fig:violin}.
The packages with high \textit{CRC} should enforce strict branch protection to reject any non-reviewed commits, or not include any non-production files in the package that have loose restrictions on code-review. On the other end of the spectrum, packages with low \textit{CRC} may be in need of more manpower. Future research may look at recommending reviewers for packages that are highly used but maintained by a small group of maintainers. 

\textbf{Post-release code vetting:} The trade-off that comes with code review is slowed-down development, while maintainers of some packages may not welcome reviewers in their projects. Further, multiple maintainers can collude in developing a package and wait until the package becomes sufficiently popular before pushing in a backdoor as part of a  long-term cyberattack. An alternate approach to code review can be the post-release crowd-vetting of the code changes in a new update~\cite{zimmermann2019small}. Package registries can provide a system where developers from all around the world can review and approve each new release of a package, while the client projects wait a certain period until an update has garnered enough approval before accepting it in their codebase.

% \subsection{Why not commit review coverage?:}
% \subsection{Malicious updates in the wild:}
% \subsection{Alpha-Omega project:}

% \section{Future work}

% \subsection{Security and Quality of code review:}
% \subsubsection{sock puppet account:}
% \subsubsection{secure reviewer hash:}
% \subsubsection{reviewer expertise:}
% \subsubsection{reviewer diversity:}

% \subsection{Reviewer recommendation:}
% \subsubsection{Recommending reviewer from the user base of a package:}

% \subsubsection{Post-release crowd vetting:}

% \subsubsection{Tool based code-review:} 

% \subsection{Provenance of generated files:}
% \subsubsection{How to check integrity of binaries, transpiled files, and other auto-generated files?:}

% \subsection{Beyond code review:}
% \subsubsection{What checks do developers want?:}
% \subsubsection{Developer perception on code review and code vetting:}
% \subsubsection{Evaluation of update security checks:}

% \subsection{Depdive Implementation:}
% \begin{itemize}
%     \item filepath matching heuristic
%     \item release commit heuristic
%     \item determining if commit was code reviewed
% \end{itemize}
\section{Related Work}
\label{relwork}

%\begin{itemize}
\textbf{Supply chain security:} Recent works have focused on the secure use of open source dependencies as part of the software supply chain~\cite{zahan2021weak, ohm2020backstabber, imtiaz2021comparative, zimmermann2019small}. Duan et al. have proposed static and dynamic analysis approaches to detect malicious packages for the interpreted languages~\cite{duan2020towards}, while Sejfia et al. have proposed machine learning models to detect malicious npm packages~\cite{sejfia2022practical}. Further, Ferreira et al. have proposed a permission-based protection mechanism for malicious npm updates~\cite{ferreira2021containing}. Our work differs from these prior works in the way that we do not explicitly aim to detect malicious updates, but rather propose an automated check for each update to aid developers in reviewing the security and the quality  of the incoming changes before accepting them into the codebase. 

\textbf{Ecosystem-wide analysis:} Recent works have done ecosystem-wide analysis to understand the state of different software supply chain networks. Zimmermann et al.~\cite{zimmermann2019small} and Liu et al.~\cite{liu2022demystifying} have looked at the vulnerability propagation in the npm ecosystem,
while Alfadel et al.~\cite{alfadel2021empirical} have investigated the PyPI ecosystem. Similarly, Zahan et al.~\cite{zahan2021weak} and Bommarito et al.~\cite{bommarito2019empirical} have looked at various quality issues in the npm and PyPI ecosystems, respectively. Further, Imtiaz et al.~\cite{imtiaz2021open} have looked at how packages release security fixes with an ecosystem-wide analysis for seven package registries.

\textbf{Code review:} There is a rich body of literature establishing the benefits of code review in software development~\cite{sadowski2018modern, mcintosh2014impact, kononenko2015investigating}. Research has shown that code review can help both software quality and security~\cite{mcintosh2016empirical, bosu2014identifying}, while also helping disseminate the knowledge of the codebase~\cite{bacchelli2013expectations}.

\section{Conclusion}
\label{conclusion}
We implement Depdive, a dependency update audit tool for Crates.io, npm, PyPI, and RubyGems packages, that (i) identifies the files and the code changes in an update that cannot be traced back to the package's source repository, i.e., \textit{phantom artifacts}; and then (ii) measures what portion of changes in the update excluding the phantom artifacts has passed through a code review process, i.e., \textit{code review coverage}. Depdive can help package users in focusing their review effort on the phantom artifacts and the non-reviewed code when pulling in a new update, while also providing a quality estimate of the incoming changes. We ran Depdive over the latest ten updates of the most downloaded 1000 packages in each of the four above-mentioned registries, of which Depdive could successfully analyze 73.6\% of the updates.

Overall, from our empirical evaluation of 24,944 updates across 3,214 packages, we present interesting insights regarding the studied package ecosystems. We find that phantom artifacts are not uncommon in the updates, either due to legitimate reasons, such as in the case of programmatically-generated files, or from presumably accidental inclusion, such as in the case of git-ignored files. While phantom artifacts are rare in Crates.io and RubyGems updates, we find that npm and PyPI updates can commonly have phantom artifacts in the form of transpiled JavaScript code, compiled binaries, and other machine-generated files.  

Regarding code review coverage, we find that updates are typically only partially code-reviewed (52.5\% of the time). Further, only 9.0\% of the packages in our dataset had all their updates fully code-reviewed, highlighting the fact that even the most used packages ship unreviewed code. We also observe that updates tend to have either very high \textit{CRC} or very low \textit{CRC}, indicating that packages at the opposite end of the spectrum require different treatments.
We further evaluated our empirical findings through a maintainer agreement survey. Overall, this paper provides an empirical evaluation of our proposed approach to auditing a dependency update and an ecosystem-level analysis of code review coverage among the latest updates of the most downloaded packages.

\bibliographystyle{IEEEtran}
\bibliography{bibliography}
\end{document}